\documentclass[twocolumn,10pt,longbibliography ]{revtex4-2}%
\usepackage[paperwidth=210mm,paperheight=297mm,centering,hmargin=2cm,vmargin=2.5cm]{geometry}
\usepackage{amsfonts}
\usepackage{amsmath}
\usepackage{amssymb}
\usepackage{bm}
\usepackage{graphicx}
\usepackage{color}
\usepackage{soul}
\usepackage{epsfig}%
\usepackage[dvipsnames]{xcolor} 
\usepackage[colorlinks=true,allcolors = black,citecolor=blue]{hyperref} 
\usepackage{natbib}
\usepackage{siunitx}

\usepackage[normalem]{ulem} % for strikeout command \sout

\begin{document}
\title{Quantum annealing sampling with a bias field}

\author{Tobias Grass}
\affiliation{ICFO - Institut de Ciencies Fotoniques, The Barcelona Institute of Science and Technology, Av. Carl Friedrich Gauss 3, 08860 Castelldefels (Barcelona), Spain}

\begin{abstract}
The presence of a bias field, encoding some information about the target state, can enhance the performance of quantum optimization methods. Here we investigate the effect of such a bias field on the outcome of quantum annealing sampling, at the example of the exact cover problem. The sampling is carried out on a \verb|D-Wave| machine, and different bias configurations are benchmarked against the unbiased sampling procedure. It is found that the biased annealing algorithm works particularly well for larger problem sizes, where the Hamming distance between bias and target configuration becomes less important. This work motivates future research efforts for finding good bias configurations, either on the quantum machine itself, or in a hybrid fashion via classical algorithms.
\end{abstract}

\maketitle

\section{Introduction}
Quantum annealing is a computational strategy for solving complex optimization problems on a programmable quantum device \cite{Albash2018,Hauke2020}. To this end, the optimization problem is formulated as a ground state search problem, by expressing its cost function as a Hamiltonian $H_{\rm P}$, typically an Ising model. Then, ground state search is carried out by producing quantum fluctuations via a driver Hamiltonian $H_{\rm drive}$, which does not commute with $H_{\rm P}$, and by reducing these fluctuations during the annealing procedure \cite{Kadowaki1998}. Two different shapes of quantum annealing can be distinguished: (i) Adiabatic quantum annealing (AQA), proposed in Ref.~\cite{Farhi2001}, demands that initially the system is prepared in the ground state of, typically, the pure driver Hamiltonian. By slowly ramping $H_{\rm drive}$ down and $H_{\rm P}$ up, the state is  adiabatically transformed into the ground state of $H_{\rm P}$. The adiabatic theorem is able to guarantee the success of this procedure, but to this end, it demands that the ramp speed is sufficiently slow (depending on the gap above the actual ground state). (ii) The other approach, quantum annealing sampling (QAS), typically performs the same actions as AQA, but at a ramp speed which invalidates the adiabatic theorem. Therefore, the annealer is not at all guaranteed to reach the ground state of  $H_{\rm P}$, but instead, it is expected to end up in a superposition of low-energy states. Then, a projective measurement at the end of the annealing process randomly selects one state out of this manifold, and repeated annealing runs sample over this manifold. The success of this strategy is heuristically based on the expectation that the ground state is still prominently represented in the final superposition.

Adiabatic state preparation, the key element of the AQA algorithm, is routinely performed in the context of quantum simulation, see for instance Ref.~\cite{steffen03} for an early realization of a small-scale quantum annealer, or Ref.~\cite{Kim2010} reporting the early implementation of the Ising model in a trapped ion quantum simulator. For the more complicated case of glassy Ising couplings, which are relevant in the context of NP-hard optimization problems \cite{barahona}, the feasibility of adiabatic state preparation in trapped ion systems has been demonstrated theoretically \cite{hauke15,grass16,raventos18}. However, several studies have argued that the scaling of the annealing gap with system size might be a prohibitive bottleneck to the AQA method \cite{joerg08,young10,joerg10,altshuler10,knysh16}. Various strategies to overcome this bottleneck have been considered in the literature: inhomogeneous driver fields \cite{dickson11,farhi11,dickson12,lanting17}, non-stoquastic driver Hamiltonians \cite{seki12,crosson14,hormozi17,albash19,oxfidan19}, tailored ramp protocols \cite{lin20,irsigler22}, reverse or biased annealing \cite{Perdomo-Ortiz2011,duan13,ramezanpour17,ohkuwa18,baldwin18,grass19,Ramezanpour2022}, as well as combinations thereof \cite{tang21}.

From the point of view of practical implementation, QAS has been more relevant than AQA. In particular, there exists a programmable quantum device which allows to carry out QAS operations with thousands of qubits, produced and commercialized by the \verb|D-Wave| company \cite{dwave11}. Unfortunately, also QAS suffers from exponentially small annealing gaps, as they suppress the weight of the ground state in the final state. Accordingly, also QAS requires strategies which are designed to suppress the population of higher levels. In contrast to AQA, the QAS method may tolerate a certain level of noise in the system. In fact, in an early experiment with the \verb|D-Wave| device \cite{Dickson2013}, it has been demonstrated that thermal fluctuations at the start of the annealing can enhance the QAS success rate. Another strategy for performance enhancement of QAS is the  non-adiabatic version of reverse annealing \cite{chancellor17}, which has recently been implemented experimentally \cite{Chancellor2021}. Although this experiment demonstrates the success of the method, it also shows that the success rates are reduced by noise. Theoretically is was shown that noise can be so harmful to reverse annealing that standard QAS produces better results \cite{Passarelli2022}. 

In the present manuscript, we consider biased annealing as another QAS strategy. In the context of AQS, this strategy has been proposed in Ref.~\cite{grass19}. It consists of adding a longitudinal bias field to the driver Hamiltonian, such that an infinitely slow switching of the Hamiltonian is still guaranteed to reach the ground state of the problem Hamiltonian. Theoretical simulations have shown, that for finite annealing times, the presence of the bias field allows for reaching the desired ground state on time scales where AQA without bias would fail, at least for small systems and for the choice of a good bias field.  In Ref.~\cite{callison21}, simulations have shown that a bias field also enhances the performance in the ultrafast limit where the annealing procedure is replaced by a sudden quench. Motivated by these theoretical studies, the present paper reports the experimental implementation of biased QAS algorithm in a \verb|D-Wave| machine. Hence, our study benchmarks the biasing algorithm under realistic conditions which includes also noisy incoherent processes.

Specifically, we have performed biased QAS for an NP-hard optimization problem (exact cover), using the same small instances (up to $N=14$ spins) as in the theoretical work from Ref.~\cite{grass19}, and extending the system size to $N=26$. While solving the classical optimization problem of this size is still no serious computational problem for classical computers, simulating the quantum annealing dynamics of such a large system would be extremely challenging on existing  classical hardware. We have used different bias fields which differ from the correct ground state by a Hamming distance of 0 to 3 spins, and in all cases, the bias field can improve the success rate considerably. This result is in line with the results obtained for biased AQA reported in Ref.~\cite{grass19}, as well as for biased quantum approximate optimization algorithm in Ref.~\cite{yu21}. It is a clear demonstration of the potential benefit due to a bias field, and it motivates future work on strategies of how to find suitable bias configurations. 

The paper is organized as follows: In Section II, we describe the nature of optimization problem (Section II A), and the biased sampling algorithm used to solve it (Section II B). In Section III, we present our results, with the most relevant information being presented in Fig.~\ref{fig:svN}, showing the scaling of the success probabilities of the biased scheme with the problem size. In Section IV, we discuss the need of future research in order to develop feasible schemes for obtaining good bias configurations.

\section{Setup}
\subsection{Optimization problem}
We study random instances of the exact cover problem. Choosing $\sigma^z_i$ as the computational basis for every spin/qubit $i$, the cost function of the problem is given as the following Ising Hamiltonian:
\begin{align}
    H_{\rm P}= \sum_{C} h_C = \sum_{C} (\sigma^z_{C(1)}+ \sigma^z_{C(2)} + \sigma^z_{C(3)} -1 )^2.
\end{align}
At this point, we keep the Hamiltonian dimensionless, but we will specify units of energy below. The problem instance is defined through the set of clauses $C$. Each clause consists of three numbers between 1 and $N$, denoted by $C(1)$, $C(2)$, $C(3)$, which choose three out of the $N$ spins. A spin configuration in which two of the three spins point upwards ($\sigma^z=+1$) and one spin points downwards ($\sigma^z=-1$) has a cost function value $h_C=0$ with respect to this clause. The configuration is said to fulfill this clause. The goal of the problem is to find the configuration which fulfills simultaneously as many clauses as possible. From randomly generated problem instances, we have selected only those instances which have exactly one configuration fulfilling all clauses, as these instances are particularly hard-to-solve, cf. Ref.~\cite{Farhi2001}. The list of used instances, together with the code of the biased annealing on the \verb|D-Wave| is provided at Ref.~\cite{github}. 

Given a set of clauses, the problem instance can be rephrased in terms of a longitudinal Ising Hamiltonian, \begin{align}
    H_{\rm P}= \sum_{i<j} J_{ij} \sigma_i^z \sigma_j^z + \sum_i h_i \sigma_i^z. 
\end{align}
When the so-defined Ising problem is implemented on the \verb|D-Wave| machine, the device's operating system \verb|Ocean| automatically scales the parameters $J_{ij}$ and $h_i$ in such a way that ensures that they remain within, and fully exploit, the accessible parameter range. On the \verb|D-Wave 2000Q| device used in this study, the parameter ranges are $J_{ij} \in [-2,1]$ and $h_i \in [-2,2]$. The corresponding frequency units are specified by the annealing schedule, see below. To encode the problem Hamiltonian on the chimera geometry of the \verb|D-Wave 2000Q| device, a minor embedding procedure is carried out \cite{Choi2008,Boothby2016}, using the function \verb|EmbeddingComposite()|, provided by the \verb|Ocean| software.

\subsection{Biased sampling algorithm}
For the solution of the optimization problem via QAS we define a homogeneous driver Hamiltonian $H_{\rm drive}=\sum_i \sigma_i^x$, which does not commute with $H_{\rm P}$, as well as a longitudinal bias Hamiltonian $H_{\rm bias}=-\sum_i \mu_i \sigma_i^z$. This bias term energetically favors the spin configuration where $\langle \sigma_i^z \rangle = {\rm sign}(\mu_i)$ for all $i$, and therefore, if $ {\rm sign}(\mu_i)$ agrees with the solution of $H_ {\rm P}$ in  all or many values of $i$, the presence such a bias Hamiltonian enhances the annealer's success rate. 

\begin{figure}
\includegraphics[width=0.48\textwidth]{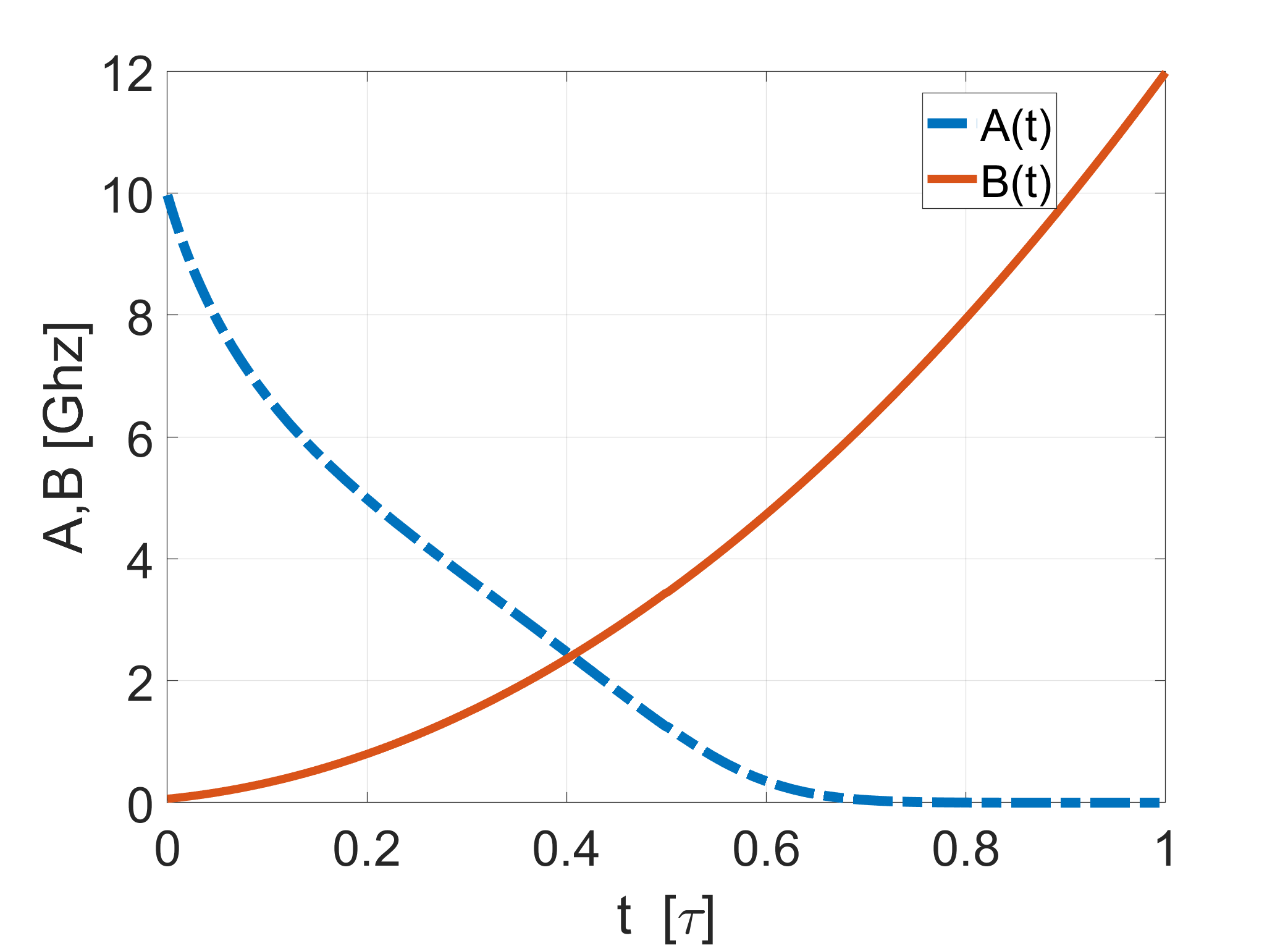}
\caption{Annealing schedule used for the sampling. \label{fig:schedule}}
\end{figure}

In the AQA version of biased annealing, as described in Ref.~\cite{grass19}, the bias field is included in the driver Hamiltonian, i.e. it is switched off during the annealing. The implementation of biased QAS, presented here, differs from the earlier work, as the bias is switched on simultaneously with the problem Hamiltonian. Accordingly, the full annealing Hamiltonian reads:
\begin{align}
    H(t) = A(t) H_{\rm drive}  + B(t) [H_{\rm P} + H_{\rm bias}]. 
\end{align}
In the beginning (at $t=0$),  $A(0) \gg B(0)$, whereas at the end of the annealing, at time $t=\tau$, $B$ has exponentially decreased to $B(\tau) \rightarrow 0$, and $A$ has grown to $A(\tau) \approx B(0)$. The explicit schedules $A(t)$ and $B(t)$ are shown in Fig.~\ref{fig:schedule}.
For the sampling, we have chosen different annealing times $\tau$ in the range between 1 $\mu$s (the fastest possible choice) and 15 $\mu$s.

From the perspective of AQA, it is an advantage to include the bias field in the driver, because with this choice, the adiabatic theorem still can guarantee that the annealer ends up in the ground of $H_{\rm P}$, even for a bad choice of $\mu_i$.  On the other hand, for the QAS method, the adiabatic theorem is irrelevant. Including the bias field into the problem part of the Hamiltonian is, first and foremost, motivated by practical considerations. Although the D-wave allows to control the time dependence of the coupling terms (i.e. the terms of type  $J_{ij} \sigma_i^z \sigma_j^z$) and the longitudinal fields (i.e. the terms of type $h_i\sigma_i^z$) independently from each other through the function \verb|h_gain_schedule|, we cannot  independently control the problem and the bias Hamiltonians, because both parts contribute to the longitudinal terms. However, from the point of view of QAS, the presence of even a bad bias term at the end of the sampling dynamics is harmless, because at that stage of the annealing schedule, the quantum fluctuations are too weak to produce significant changes anymore. What matters instead is to have, at intermediate times before the sampler starts freezing, the combined presence of all three terms $H_{\rm drive}$, $H_{\rm bias}$, and $H_{\rm P}$. To this end, it does not matter whether $(H_{\rm drive} + H_{\rm bias})$ or  $(H_{\rm bias}  + H_{\rm P})$ are controlled jointly.

\section{Results}
\begin{figure}[t]
    \centering
    \includegraphics[width=0.48\textwidth]{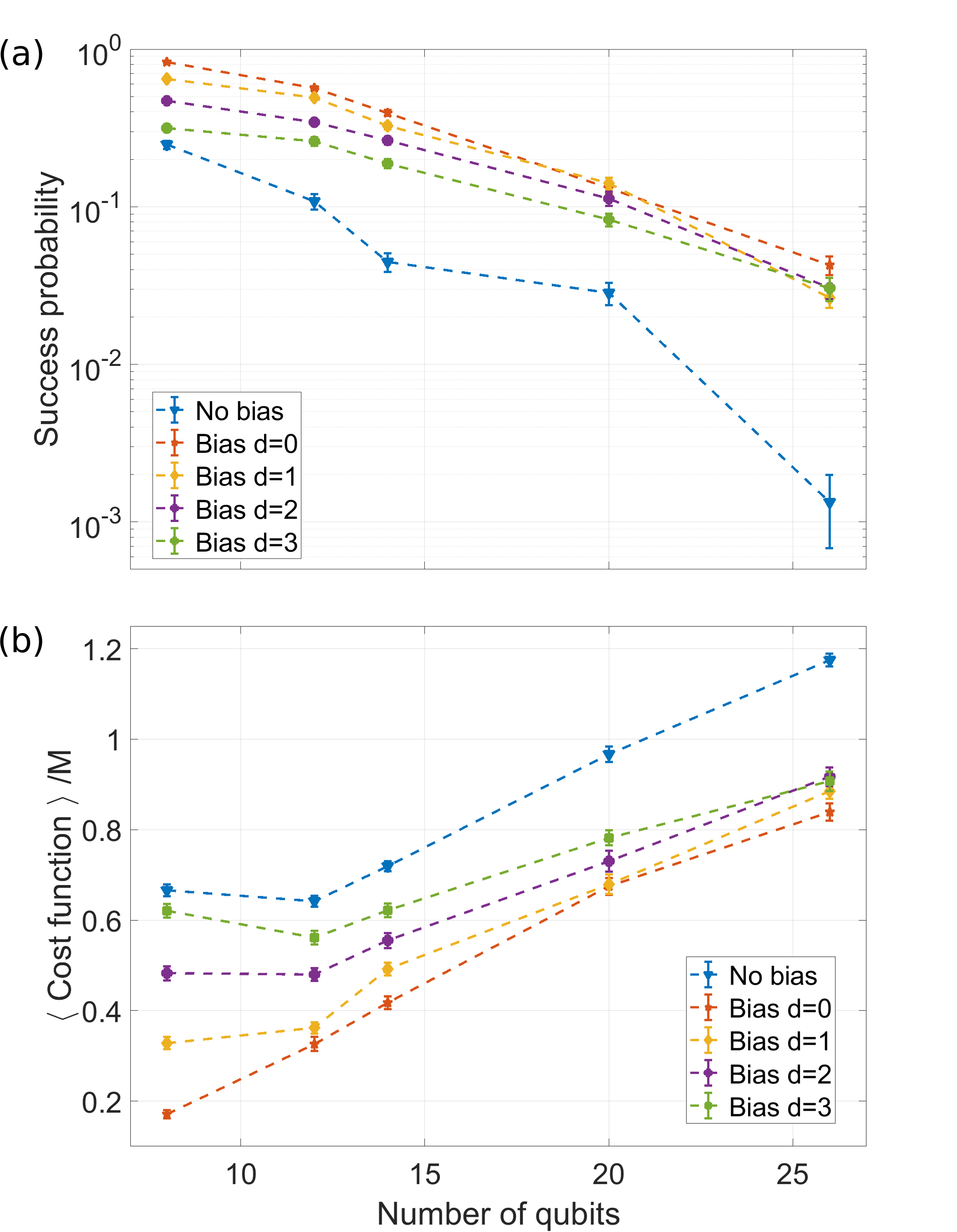}
    \caption{The performance of biased QAS is benchmarked and compared to unbiased QAS, by considering success probability and average cost function value at the end of the anneal (divided by the number $M$ of clauses) as a function of system size, and for different biases, distinguished by their Hamming distance $d$ to the optimal solution. Here, the annealing time $\tau=1 \mu$s is fixed. The error bars represent the standard error of the mean.}
    \label{fig:svN}
\end{figure}

We have performed QAS for the exact cover problem, as specified in the previous section, for problem instances denoted $(N,\alpha)$, where $N$ defines the number of spins/qubits, and $\alpha$ is an index for each of the random instances. We consider sizes between $N=8$ to $N=26$, and for each size, we consider $N_{\alpha}=100$ random instances, that is, $\alpha\in[1,100]$. From the solution of the problem Hamiltonian, which in our case are \textit{a priori} known and which we denote by $\{s_{i}\}$, we have constructed bias fields $\{\mu_i\}$, such that the Hamming distance $d=\sum_i |\mu_i - s_i|$ between bias and target is fixed. For every instance and for each value of $d$, we perform $N_{\rm anneal}=30$ anneals, and we count the number of times $N_{\rm success}$ in which the final measurement outcome agrees with the target configuration. Then, the corresponding success probability $p_{(N,\alpha)}(d)=N_{\rm success}/N_{\rm anneal}$ is averaged over all instances of a given size, $p_N(d)=\sum_\alpha p_{(N,\alpha)}(d)/N_{\alpha}$, and the result is plotted in Fig.~\ref{fig:svN}(a), with error bars representing the standard error of the mean. We also determine, after every anneal, the cost function corresponding to the obtained configuration. The average over all anneals and all instances, as a function of system size $N$ and for different bias choices $d$, is shown in Fig.~\ref{fig:svN}(b). For the annealing time $\tau$ in Fig.~\ref{fig:svN}, we have used the fastest possible choice, $\tau=1\mu$s. For comparison, we have also performed QAS without bias field.

As expected, the obtained data provides evidence that any of the chosen bias fields enhances the performance of the annealer. Importantly, the enhancement becomes more significant as the system size is increased. To some extent, this behavior is also expected from the study of the biased AQA \cite{grass19}, and can be attributed to the fact that, at a fixed Hamming distance $d$,  the error ratio of the bias, $r \equiv d/(N-d)$, that is, the ratio of ``bad'' to ``good'' bias terms, decreases with $N$. However, let us compare, for instance, the results at $N=8$ and $d=1$ with the results at $N=26$ and $d=3$: Although in these two configurations, the bias fields have a similar error ratio $r$, the enhancement factor (defined as success probability with bias divided by success probability without bias) turns out to be quite different: At $N=8$, the enhancement is by a factor of $2.6\pm0.2$, whereas at $N=26$, we obtain enhancement by a a factor of $23\pm16$. Despite the large statistical uncertainty in the latter number, owed to very low success rate of the unbiased annealing at $N=26$, these numbers suggest that the biased QAS method is particularly strong for larger problem instances. To further appreciate this strength of the biased method, it is also illustrative to look at the number of instances which, after 30 anneals, have been solved correctly. At $N=26$ and $\tau=1 \mu$s, there are only 4 (out of 100) instances which have been solved without bias. This shows that unbiased QAS is essentially not able to find the solution for this problem size at this short annealing time. On the other hand, even a $d=3$ bias elevates the percentage of correctly solved instances to almost 50 percent (47 out of 100 instances). Another interesting observation which can be made for large system sizes is the fact that, within the studied range of $d$, the Hamming distance $d$ of the bias affects the annealing outcome only weakly. While at small $N$, both success probability and average cost function appear ordered according to $d$, for $N=26$ the bias with $d=1$ achieves a slightly lower success probability than for biases with $d=2$ and $d=3$. As seen from the error bars in the plot,  this ``under''-performance can be attributed to statistical fluctuations, and in fact at large problem sizes, all the different bias choices perform similarly well.

\begin{figure}[t]
    \centering
    \includegraphics[width=0.48\textwidth]{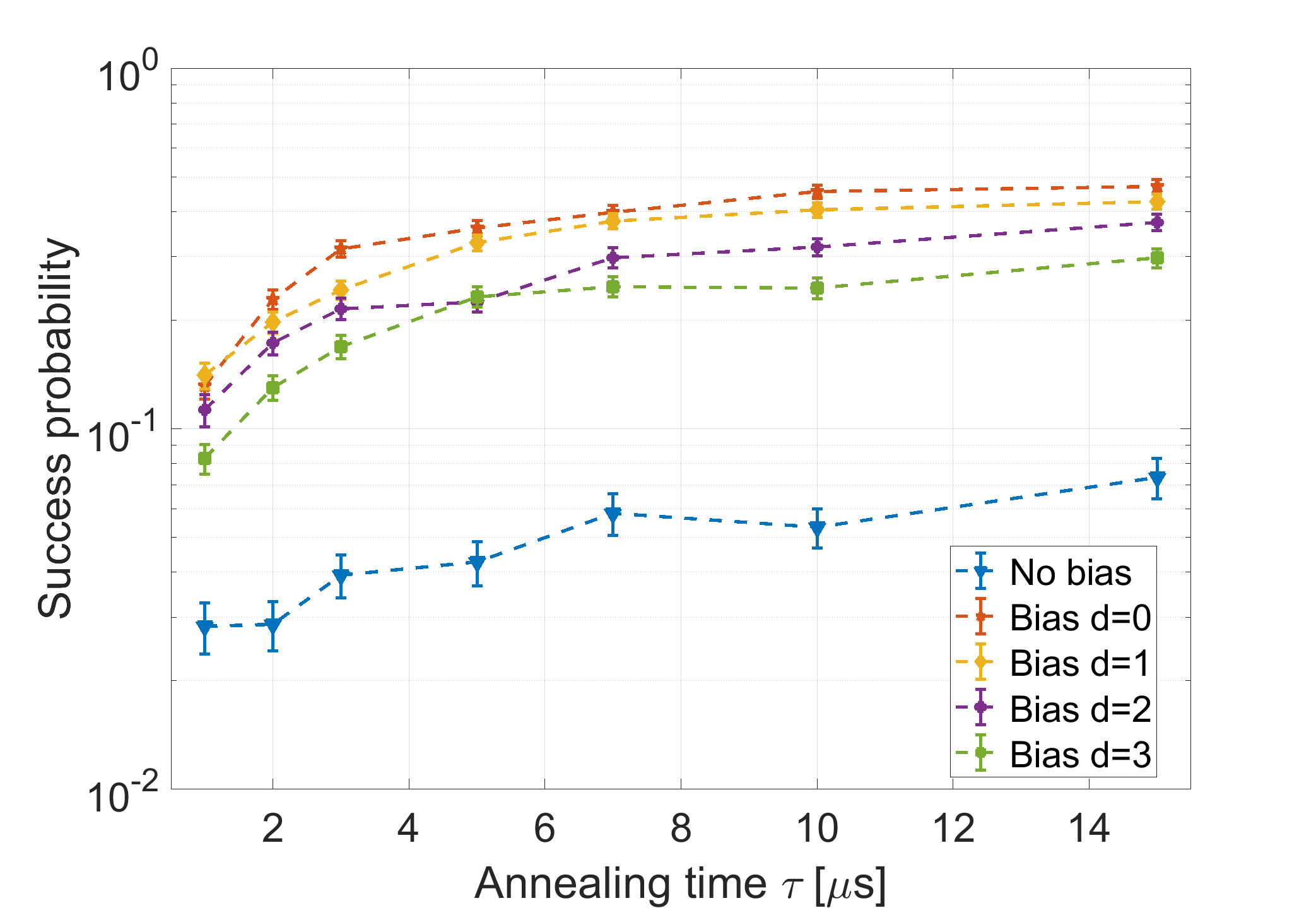}
    \caption{We investigate the dependence of biased and unbiased annealing on the annealing time $\tau$. We focus on $N=20$ qubits, and plot the averaged success rates for unbiased QAS and biased QAS with different choices of the bias' Hamming distance $d$ from the correct solution. The error bars represent the standard error of the mean.}
    \label{fig:svt}
\end{figure}

\begin{figure}[t]
    \centering
    \includegraphics[width=0.48\textwidth]{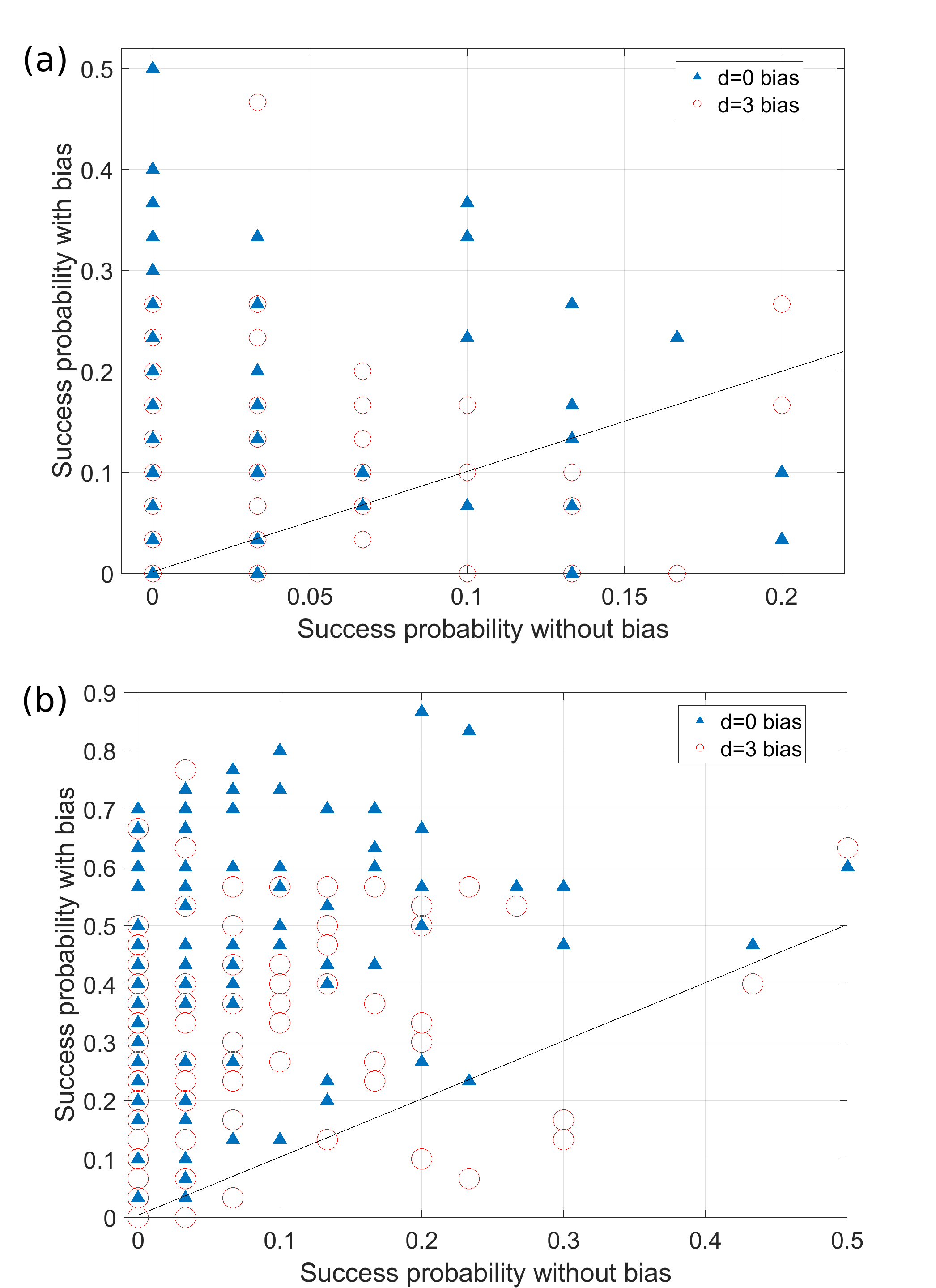}
    \caption{We plot the success probabilities of biased QAS (with $d=0$ and $d=3$) vs. the success probabilities of unbiased QAS, for individual instances with $N=20$ qubits. In (a), the annealing time $\tau=1\mu$s; in (b), the annealing time $\tau=15\mu$s. We note that, owed to the fact that the success probabilities are calculated from $N_{\rm anneal}=30$ annealing runs, there is a limited discrete set of possible values, and hence it occurs that several (of the in total 100) instances have identical success probabilities, that is, that the corresponding data points cannot discerned from each other.}
    \label{fig:pvsp}
\end{figure}

In Fig.~\ref{fig:svt}, we study the dependence of the success probability on the annealing time $\tau$ for $N=20$. Obviously, increasing $\tau$ increases the success probability, both with and without bias fields. This improvement is seen to happen at similar rates for different bias configurations and in the unbiased case. For instance, the success probability for the $d=3$ bias increases by a factor $3. 6\pm0.5$, when increasing $\tau$ from 1$\mu$s to 15$\mu$s, and by factor $2.6\pm0.8$ for unbiased QAS. From this, we conclude that the choice of $\tau$ does not seem to be crucial for benchmarking biased against unbiased QAS via average success probabilities. A more differentiate picture can be obtained when analyzing the success probability of individual instances. In Fig.~\ref{fig:pvsp}, we present a comparison between the success probabilities of biased and unbiased annealing for individual instances. Again, we concentrate on $N=20$, and show the data for $\tau=1\mu$s, as well as for $\tau=15\mu$s, considering bias fields with $d=0$ and $d=3$. It is obvious that in all cases the majority of instances profit from the presence of a bias field, but the number of instances which do not take advantage from the bias is significantly reduced by increasing $\tau$. For instance, at $\tau=1\mu$s, only 78 (65) instances (out of 100) have higher success probabilities in biased QAS with $d=0$ ($d=3$) than in unbiased QAS. In the case of $\tau=15\mu$s, however, this is the case for 98 (89) instances. From this perspective, the advantage of a bias field appears to be increased if the annealing is performed at lower speed. 

We also note that, if the annealing time $\tau$ was the only time scale in the annealing protocol, increasing $\tau$ would, for all $N$, increase the average time-to-solution, because the best ratios between average success probability over $\tau$ are achieved at $\tau=1 \mu$s. However, since most of the computational time is actually spent by the programming of the annealer rather than by the annealing procedure itself, the best choice of $\tau$ is not obvious from these numbers alone, and it can actually be beneficial to choose a larger value of $\tau$.

\section{Discussion and Outlook}
For producing the improvements via bias fields which have been demonstrated in the previous Section, strategies to choose good bias fields are needed. This choice requires some knowledge about the target state, which here we have assumed to be given \textit{a priori}. In practice, though, this might not be the case. A strategy to find good bias configurations had been proposed in Ref.~\cite{grass19}: The outcome of a previous annealing run can be used as a bias in the subsequent anneal, which can iteratively improve the bias, until the correct solution is found. One limitation of this strategy, discussed in the context of theoretical simulations of the algorithm in Ref.~\cite{grass19}, is the fact that in certain cases the initially produced bias might actually attract the annealer towards an excited state rather than the ground state. In a practical implementation, however, the iterative scheme suffers from another, maybe more significant drawback: As already mentioned above, the time $\tau$ (or even $N_{\rm anneal}\tau$) is typically smaller than the time which is required for the programming of the \verb|D-Wave| device. Since in the iterative procedure, the annealer has to re-programmed  after every anneal (or after some sampling period), the time efficiency of iterative QAS becomes very low. Possibly, in future hardware implementations, the fact that the iterative updates occur only in the magnetic field terms but not in the couplings might be exploited to speed up iterative schemes.

While the iterative scheme is appealing because it can be carried out entirely on a quantum device, future research shall also explore the possibility of determining good bias fields through classical algorithms, e.g. through variations of the simulated annealing algorithm \cite{kirkpatrick83}. A final assessment whether biased annealing has a real advantage to unbiased annealing can only be made if the computational cost of finding a suited bias is taken into account, but the improvements due to bias fields, seen in the present work, appear to be very promising.

\begin{acknowledgments}
The project that gave rise to these results received the support of a fellowship from ``La Caixa'' Foundation (ID100010434). The fellowship code is LCF/BQ/PI19/11690013. I also acknowledge support from ERC AdG NOQIA; Agencia Estatal de Investigación (R\&D project CEX2019-000910-S, funded by MCIN/
AEI/10.13039/501100011033, Plan National FIDEUA PID2019-106901GB-I00, FPI,
QUANTERA MAQS PCI2019-111828-2, Proyectos de I+D+I “Retos Colaboración”
QUSPIN RTC2019-007196-7), MCIN via European Union NextGenerationEU (PRTR-
C17.I1); Fundació Cellex; Fundació Mir-Puig; Generalitat de Catalunya through the European Social Fund FEDER and CERCA program (AGAUR Grant No. 2017 SGR 134, QuantumCAT \ U16-011424, co-funded by ERDF Operational Program of Catalonia 2014-2020); EU Horizon 2020 FET-OPEN OPTOlogic (Grant No 899794); National Science Centre, Poland (Symfonia Grant No. 2016/20/W/ST4/00314); European Union’s Horizon 2020 research and innovation programme under the Marie-Skłodowska-Curie grant agreement No 101029393 (STREDCH) and No 847648 (“La Caixa” Junior Leaders
fellowships ID100010434: LCF/BQ/PI20/11760031, LCF/BQ/PR20/11770012, LCF/BQ/PR21/11840013).
\end{acknowledgments}

\bibliography{lib}
\end{document}